\documentstyle[preprint,aps]{revtex}

\tighten

\begin{document}

% \draft

\title{String breaking by dynamical fermions in three-dimensional lattice QCD}

\author{Howard D. Trottier}
\address{Newman Laboratory of Nuclear Studies, Cornell University,
Ithaca, NY 14853-5001, and \\ Physics Department, Simon Fraser
University, Burnaby, B.C., Canada V5A 1S6\cite{Permanent}}

\maketitle

\begin{abstract}
The first observation is made of hadronic string breaking due to
dynamical fermions in zero temperature lattice QCD. The
simulations are done for SU(2) color in three dimensions, with two
flavors of staggered fermions. The results have clear implications
for the large scale simulations that are being done to search (so
far, without success) for string breaking in four-dimensional QCD.
In particular, string breaking is readily observed using only
Wilson loops to excite a static quark-antiquark pair. Improved
actions on coarse lattices are used, providing an extremely
efficient means to access the quark separations and propagation
times at which string breaking occurs.
\end{abstract}

\newpage
\section*{}

The string picture of quark confinement predates Quantum
Chromodynamics, and continues to play a central role in
theoretical efforts to characterize the physics of QCD
\cite{Kuti}. Simulations of quenched lattice QCD have in fact
demonstrated that color-electric field lines connecting a static
quark and antiquark are squeezed into a narrow tube
\cite{Michael}, in accord with many models of the linearly rising
static potential.

In QCD with dynamical fermions the flux-tube joining static quarks
is expected to be unstable against fission at large $R$, where
there is enough energy in the fields to materialize light quarks
from the vacuum, which bind to the heavy quarks to form a pair of
color-neutral bound states. The simplest indicator of string
breaking is provided by the static quark-antiquark potential
$V(R)$, which should approach a constant at large $R$
\begin{equation}
   V(R\to\infty) = 2 M_{Q\bar q} ,
\label{Vbreak}
\end{equation}
where $M_{Q\bar q}$ is the mass of a bound state of a heavy quark
and a light antiquark.

There is as yet no convincing evidence of string breaking in
lattice simulations at zero temperature, despite extensive
simulations by several large scale collaborations
\cite{Kuti,CPPACS,UKQCD,SESAMTchiL,DeTar}. A recent conjecture
\cite{Gusken,Drummond}, which has received widespread attention,
holds that the failure to detect the broken string state is due to
a very poor overlap of the Wilson loop operator with the true
ground state of the system at large $R$ (Wilson loops have been
used almost exclusively to construct trial states in these
simulations). The possibility has been raised that the broken
string state may even be unobservable with Wilson loops for all
practical purposes \cite{CPPACS}. This picture has recently
stimulated interest in simulations using other operators to excite 
the system \cite{Wittig,Koniuk}.

Evidence for a somewhat different point of view is presented in
this paper. The task of observing string breaking in full QCD is
computationally very challenging, given the high cost of
simulating dynamical fermions. These simulations have so far been
done on lattices with a relatively fine grid. Consequently, only a
fairly limited range of quark separations $R$ has been explored
and, more importantly, the propagation of the trial states created
by Wilson loops has only been studied for rather short Euclidean
times $T$ (e.g., $T=0.4$--$0.8$~fm in Ref. \cite{CPPACS}). The
restriction to such short propagation times precludes a definitive
assessment of whether string breaking occurs in these states. This
limitation may be substantially alleviated by working on coarser
lattices with improved actions.

This proposal is assessed here by doing simulations in
three-dimensional lattice QCD (QCD$_3$) with two flavors of
dynamical staggered fermions. QCD$_3$ provides an excellent
laboratory for studying the physics of realistic QCD (for a review
see Ref. \cite{Teper}). Quenched QCD$_3$ has been shown to exhibit
linear confinement, flux-tube formation, a deconfining phase
transition, and a rich glueball spectrum, and spontaneous
breakdown of a ``chiral'' symmetry has been observed in both the
quenched and unquenched theories \cite{Kogut}.

The first observation of string breaking by dynamical fermions is
reported here, using only Wilson loop operators to excite the
static quark-antiquark system \cite{Prelim}. Although these simulations 
are done in QCD$_3$, the results have clear implications for work on the
four-dimensional theory. In particular the simulations are done on
coarse lattices, using an improved gluon action that is accurate
through $O(a^2)$, and a staggered fermion action that is accurate
through $O(a)$. This allows the system to be much more efficiently
probed at the physical separations and propagation times relevant
to string breaking. Moreover, the overlap of an operator with the
lowest-lying state generally increases with the coarseness of the
lattice, due to the suppression of higher momentum modes in the trial 
state. In contrast, recent studies of QCD$_3$ with dynamical scalar 
matter fields \cite{Wittig}, and adjoint matter fields \cite{Poulis},
using unimproved actions on lattices with about half the spacing 
considered here, did not resolve string breaking in Wilson loops, 
presumably because insufficient propagation times were attained.

Using the quenched string tension to set the ``physical'' length
scale for QCD$_3$, these results suggest that string breaking
occurs in Wilson loops at modest propagation times, somewhat in
excess of 1~fm (on modestly coarse lattices), which is not much
beyond what has been reached in four dimensions. The physical
quark separation $R\approx1.5$~fm at which string breaking is
found in QCD$_3$ is consistent with expectations in four
dimensions \cite{Sommer}. Lattice spacings of about 0.2--0.3~fm
are used here; these lattices are about twice as coarse as those
that have been used in four-dimensional studies of string
breaking. Comparable lattice spacings could be reliably accessed
in realistic QCD using improved actions \cite{Alford}, resulting
in an enormous improvement in computational efficiency.

To begin with, consider the pure gauge theory in three dimensions.
To reduce the computational cost SU(2) color is considered.
Simulations were done using a tree-level $O(a^2)$-accurate gluon
action, allowing for different spatial and temporal lattices
spacings, $a_s$ and $a_t$ respectively \cite{Symanzik}
\begin{equation}
   {\cal S}_{\rm imp} = - \beta \! \sum_{x,\mu>\nu} \!
   \xi_{\mu\nu} \left[
   \case{5}{3} P_{\mu\nu}
 - \case{1}{12} (R_{\mu\nu} + R_{\nu\mu})
   \right] .
\label{Simp}
\end{equation}
$P_{\mu\nu}$ is one-half the trace of the $1\times1$ plaquette and
$R_{\mu\nu}$ is one-half the trace of the $2\times1$ rectangle in
the $\mu\times\nu$~plane. The bare QCD coupling $g_0^2$ (of
dimension $1$) enters through the dimensionless parameter $\beta =
4/g_0^2a_s$, and the bare lattice anisotropy is input through
$\xi_{3i} = \xi_{i3} = a_s/a_t$ and $\xi_{ij} = a_t/a_s$
($i,j=1,2$).

QCD$_3$ is a super-renormalizable theory; in fact the bare
coupling $g_0^2$ and bare quark masses $m_0$ remain finite as the
ultraviolet regulator is removed. Since $g_0^2$ and $m_0$ are
cutoff-independent in the continuum limit, one obtains extremely
simple scaling laws for physical quantities in lattice QCD$_3$. In
particular, masses in lattice units (including the input bare
quark masses) should satisfy
\begin{equation}
   \beta a m = \mbox{constant}
\label{scaling}
\end{equation}
in the continuum limit. The improved action $S_{\rm imp}$ differs
from the continuum theory at the tree-level by terms of $O(a^4)$;
one-loop corrections induce errors of $O(g_0^2 a^3)$.

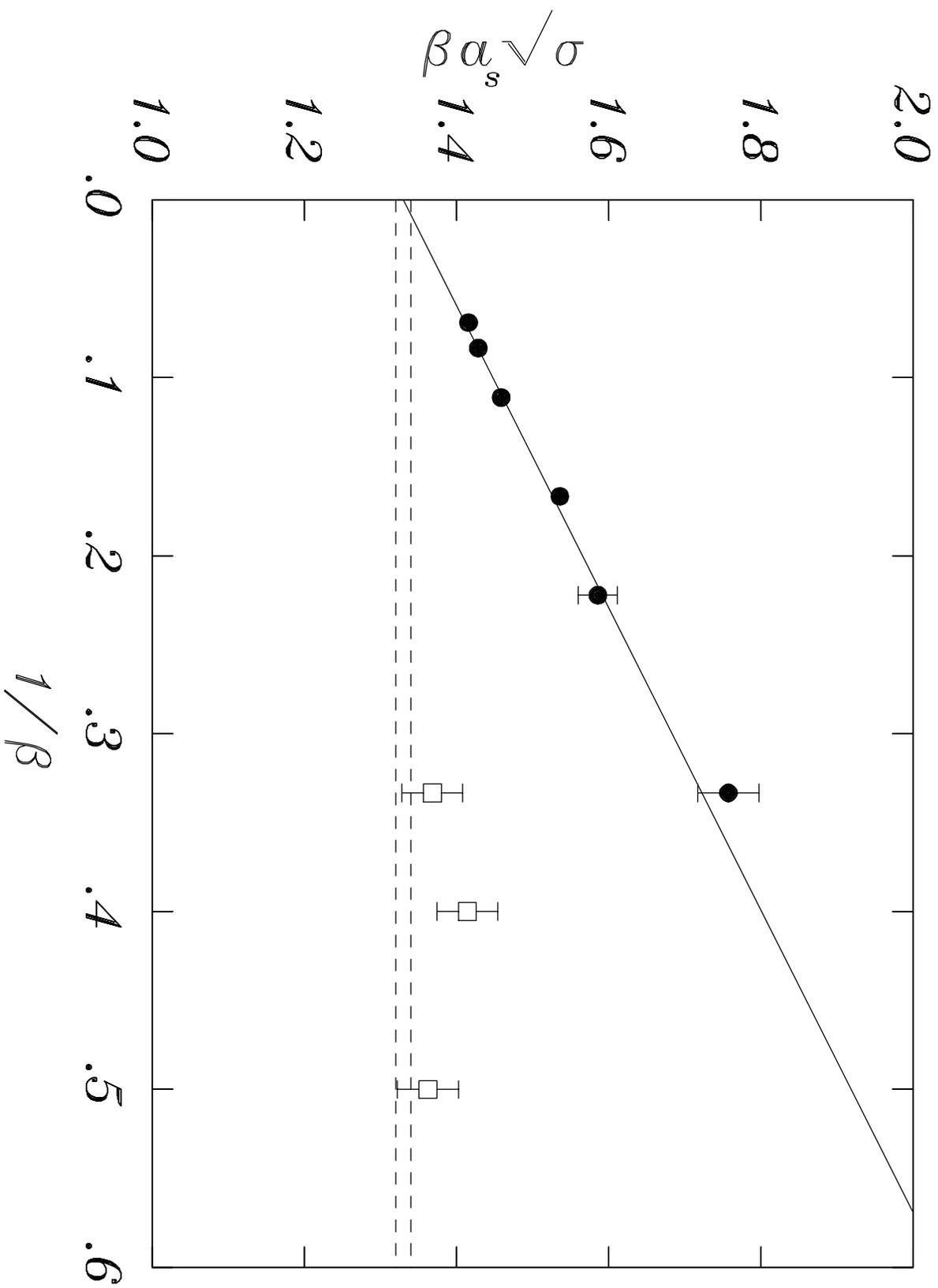
\begin{figure}[htb]
\caption{Lattice spacing dependence of the quenched
string tension from the improved gluon action ($\Box$); results
for the Wilson gluon action ($\bullet$) are taken from Ref.
\protect\cite{Teper}.} \label{fig:sigma}
\end{figure}

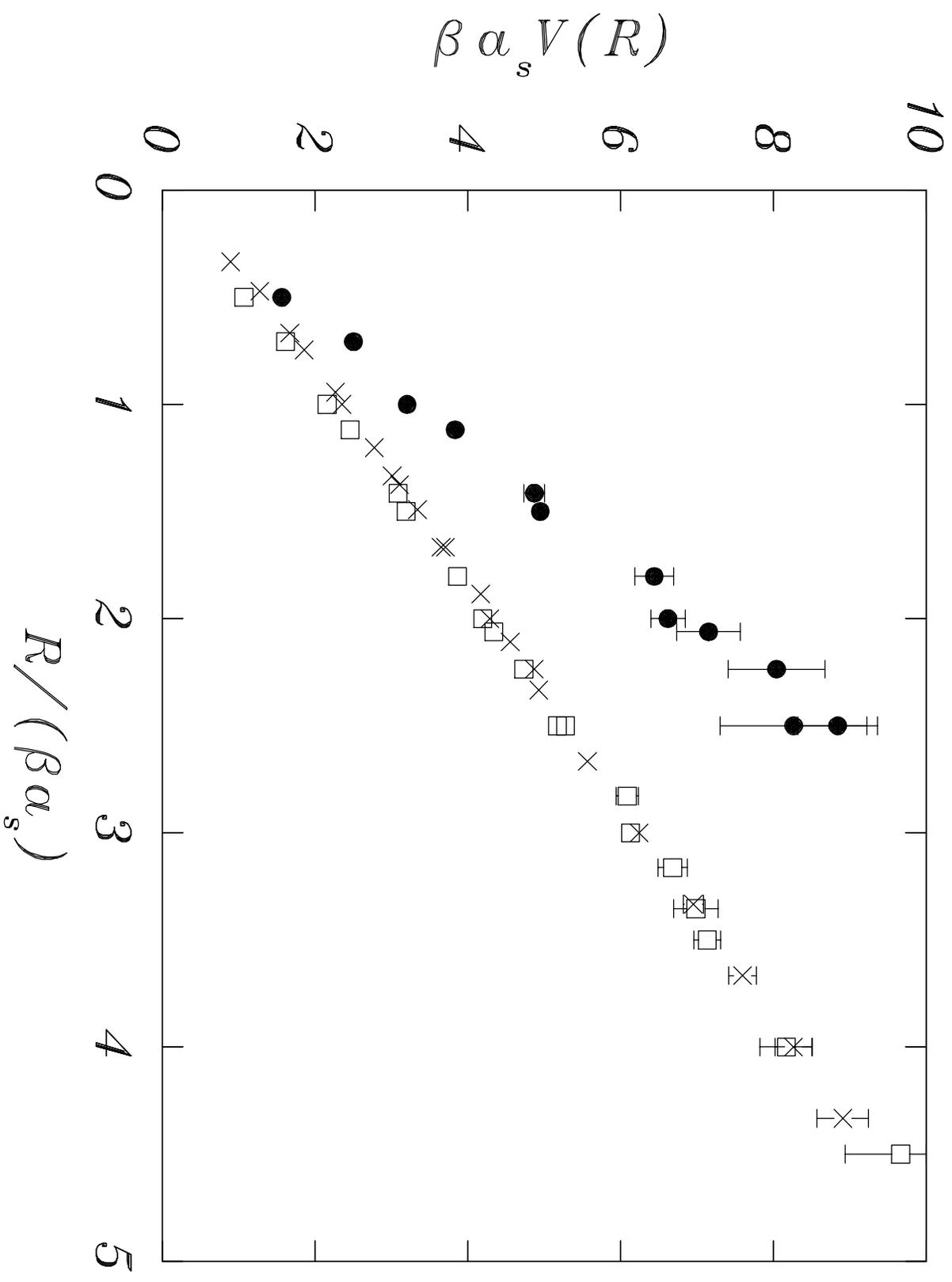
\begin{figure}[htb]
\caption{Static potential from the Wilson gluon
action at $\beta=2$ ($\bullet$), and from the improved gluon
action at $\beta=2$ ($\Box$) and $\beta=3$ ($\times$).}
\label{fig:qpot}
\end{figure}

Results for the quenched string tension obtained with $S_{\rm
imp}$ are shown in Fig. \ref{fig:sigma}; for comparison, results
obtained by Teper \cite{Teper} for the unimproved Wilson gluon
action are also shown (the dashed lines show the continuum limit
estimated in \cite{Teper}). A comparison of the potential computed
from the two actions on lattices of comparable spacing is shown in
Fig. \ref{fig:qpot}.

Simulations of $S_{\rm imp}$ were done on lattices with $\beta=2$,
2.5, and 3, all with anisotropy $a_t/a_s=1/4$. The lattice volumes
range from $16^3$ at $\beta=2$ to $24^3$ at $\beta=3$. Hybrid
molecular dynamics were used to generate the configurations (the
$\Phi$ algorithm \cite{Gottlieb} was used to
generate the unquenched configurations analyzed below); 50
molecular dynamics steps were taken with step size $\Delta t =
0.02$. A standard fuzzing procedure was employed on the link
variables used in Wilson loop measurements. Integrated
autocorrelation times satisfied $\tau_{\rm int} \lesssim 0.5$,
with typically 5--10 trajectories skipped between measurements.

A superficially surprising aspect of these results is the $O(a)$
scaling violation evident in the Wilson action data for
$\sqrt\sigma / g_0^2$. In fact this demonstrates the need to
renormalize the bare coupling at finite $a$ \cite{Lepage}. In four
dimensions it is known that perturbative expansions in the bare
lattice coupling are spoiled by large renormalizations. A
renormalized coupling defined by a physical quantity absorbs these
large corrections, and greatly improves perturbation theory for
many quantities \cite{LepMac}.

In lattice QCD$_3$ one expects to find an $O(g_0^2 a)$
renormalization of the bare coupling at one-loop order. This
produces the linear scaling violation in the unimproved results
for $\sqrt\sigma / g_0^2$. On the other hand this effect can be
removed by using a physical quantity to define a renormalized
scale. For example, Teper has done simulations of
three-dimensional glueball masses $m_G$ \cite{Teper}, and showed
that the ratios $m_G / \sqrt\sigma$ exhibit $O(a^2)$ scaling
violations.

The distinctive signature of the renormalization of the bare
coupling in QCD$_3$ exposes an interesting feature of the action.
One might expect to see a reduction in the renormalization of
$g_0^2$ when $S_{\rm imp}$ is used, compared to simulations with
the Wilson action. Remarkably one finds that the renormalization
is in fact almost completely eliminated; as seen in Fig.
\ref{fig:sigma} the error in $\sqrt\sigma / g_0^2$ at $\beta=2$ is
reduced from about 45\% with the Wilson action, to less than about
5\% with $S_{\rm imp}$. This is also apparent from the data in
Fig. \ref{fig:qpot}, where the potential $V$ and separation $R$
from the lattices at $\beta=2$ and 3 are compared on a common
scale, set by $g_0^2$. This is a genuinely surprising result,
since all of the higher dimension operators present in $S_{\rm
imp}$ should contribute to a leading $O(g_0^2 a)$ renormalization
of the bare coupling. Apparently the operator series in this
effective action converges very rapidly, even at scales near the
lattice cutoff.

One can therefore conveniently measure physical quantities
directly in terms of the bare coupling when the improved action
Eq. (\ref{Simp}) is used. It is important to note however that the
use of $S_{\rm imp}$ would prove advantageous even if $g_0^2$
required renormalization; one would simply use a measured quantity
(such as the quenched string tension) to set the scale for other
observables, and the leading $O(a^2)$ errors in the Wilson action
would thus be removed.

Simulations with dynamical Kogut-Susskind fermions were done to
look for effects of string breaking. The staggered fermion action
${\cal S}_{\rm KS}$ in three dimensions \cite{Burden} is identical
in form to the four-dimensional theory, and describes two flavors
of four-component spinors
\begin{eqnarray}
    {\cal S}_{\rm KS} = \sum_{x,\mu} \eta_\mu(x)
   \bar\chi(x) \bigl[ U_\mu(x) \chi(x+\hat\mu) \ \ \ \ \ \ & &
\nonumber \\
   - U^\dagger_\mu(x-\hat\mu) \chi(x-\hat\mu) \bigr]
   + 2 am_0 \sum_x \bar\chi(x) \chi(x) , & &
\label{Sstagg}
\end{eqnarray}
where $\eta_\mu(x) = (-1)^{x_1 + \ldots + x_{\mu-1}}$ is the usual
staggered fermion phase. Unquenched simulations were done with
improved glue on isotropic lattices ($a_t = a_s$)
at $\beta=2$ and 3, on $12^2\times8$ and $16^2\times10$ 
volumes respectively \cite{Temp}. Isotropic lattices for the 
unquenched simulations were used in order to probe the longest 
propagation times $T$ possible, for the least computational cost 
(although measurements over a finer range of $T$ on anisotropic
lattices would be advatangeous for fitting and possible extraction
of excited state energies). The input bare quark mass in 
lattice units was scaled according to
$m_0/g_0^2 = 0.075$ [cf. Eq. (\ref{scaling})], with $m_\pi /
m_\rho$ found to be $\approx 0.75$. Approximately 6,000
measurements were done at $\beta=2$, and 2,000 at $\beta=3$.

The leading discretization errors in the action Eq. (\ref{Sstagg})
are of $O(a^2)$ \cite{ImpStagg}; at finite lattice spacing the
bare quark mass must also absorb an $O(g_0^2 a)$ renormalization.
However these effects appear to induce little scaling violation
in the static potential at the small quark mass used here, perhaps
because the energy in the color fields dominates the string
breaking process (significant scaling violations in the unquenched
potential were seen when the Wilson gluon action was used).

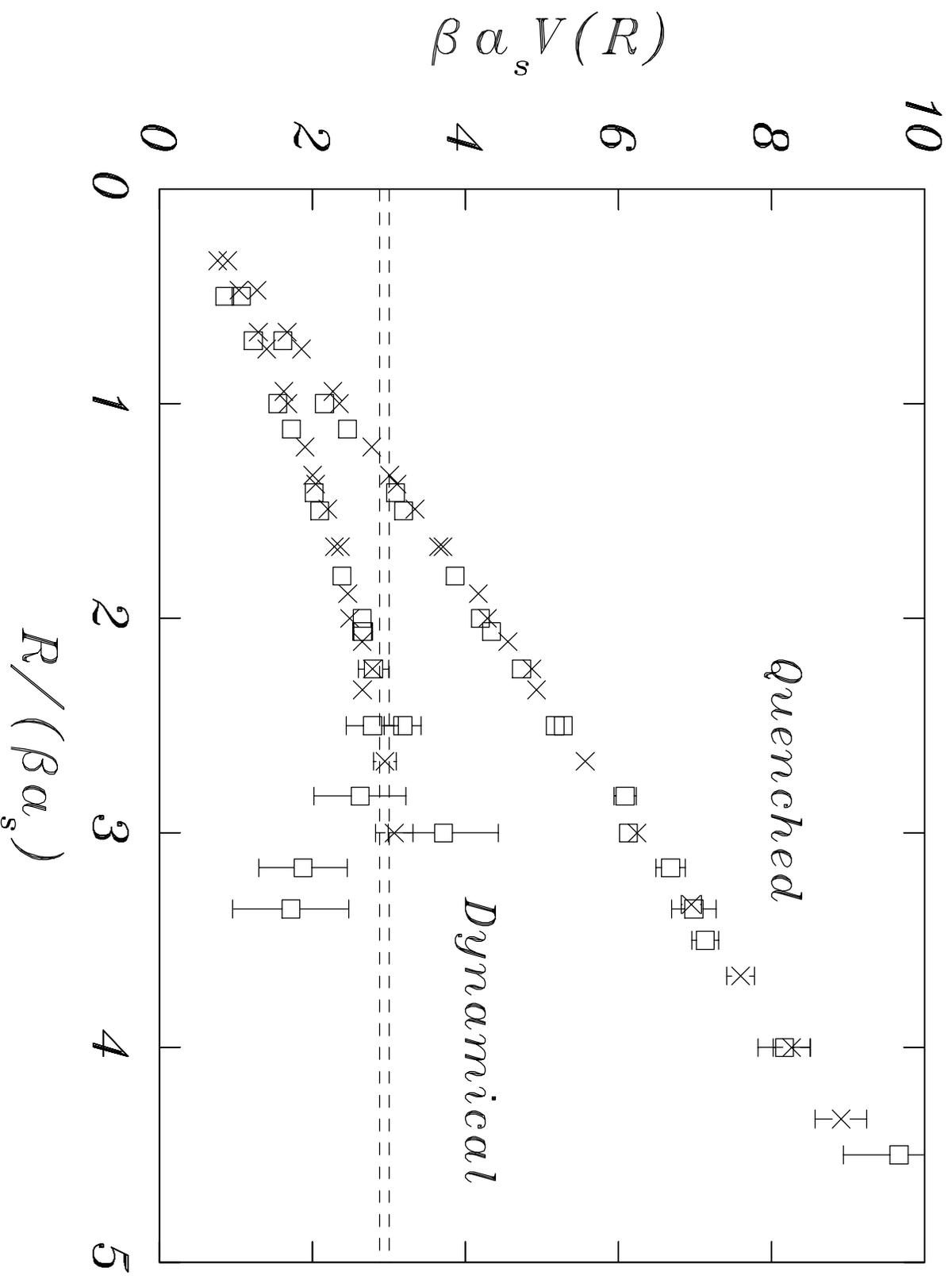
\begin{figure}[htb]
\caption{Static potential from the improved gluon
action with and without dynamical staggered fermions, at $\beta=2$
($\Box$) and $\beta=3$ ($\times$). The unquenched simulations
were done on isotropic lattices ($a_t = a_s$). 
The dashed lines are one sigma limits for 
$2\beta a_s M_{Q\bar q}$, computed on dynamical
configurations at $\beta=3$.} \label{fig:dpot}
\end{figure}

Results for the unquenched potential are shown in Fig.
\ref{fig:dpot}. The onset of string breaking is assessed by
comparing the potential energy with the mass of the two $Q\bar q$
bound states into which the system should hadronize. The staggered
heavy-light meson propagator \cite{Fiebig} was computed in the
unquenched configurations at $\beta=3$, for equal valence and sea
quark masses; the result is shown as the dashed lines in Fig.
\ref{fig:dpot}. String breaking is clearly demonstrated, with the
unquenched potential approaching the expected asymptotic value
[cf. Eq. (\ref{Vbreak})]. Excellent scaling behavior is observed;
note that the lattice spacing increases by 50\% from $\beta=3$ to
$\beta=2$. A continuum extrapolation of the string breaking
distance $R_b$ can be estimated from these results, $R_b / (\beta
a) \approx 2.5$.

The onset of string breaking is made particularly evident by a
comparison of the quenched and unquenched potentials at the same
$\beta$ (note that no adjustment of the energy zero has been made
in any of these results). The meaningfulness of this comparison is
supported by the fact that the bare coupling $g_0^2$ undergoes
little renormalization; moreover quenching effects at a given
$\beta$ were found to change the $\rho$-meson mass by less than
10\%.

Only (fuzzy) Wilson loop operators were used in these
calculations. As discussed above, little evidence of string
breaking in Wilson loops has been found in previous simulations
with dynamical fermions. A distinguishing feature of the
simulations done here, compared with earlier work, is the relative
coarseness of the lattices that were used (of course the physics
of the string breaking process could be substantially different in
three dimensions; on the other hand, simulations of QCD$_3$ on
much finer lattices \cite{Wittig,Poulis} did not resolve string 
breaking in Wilson loops).

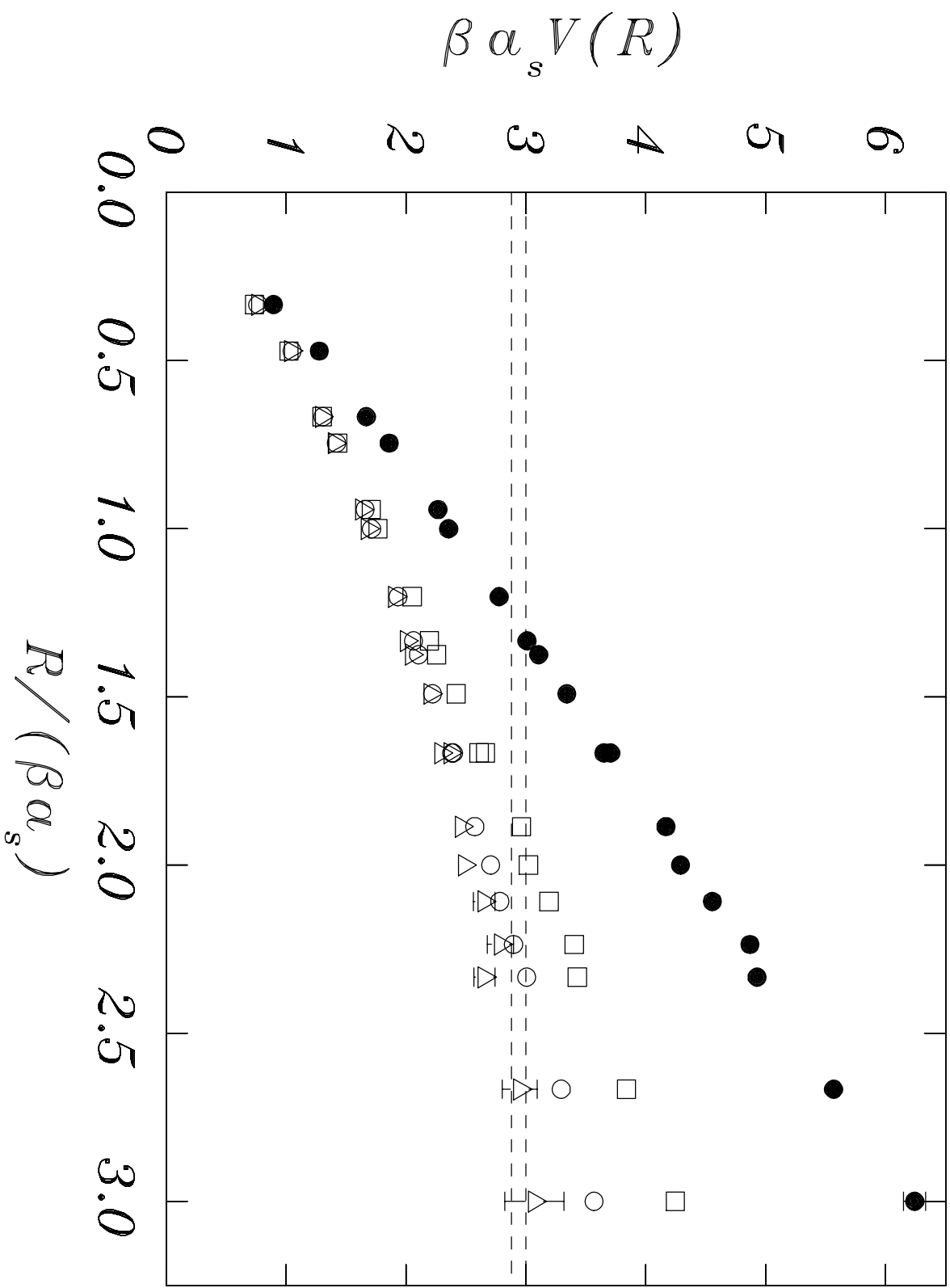
\begin{figure}[htb]
\caption{Time-dependent effective potential $V(R,T)$
versus $R$, in the dynamical configurations at $\beta=3$, for
$T/a_s=1$ ($\Box$), $3$ ($\circ$), and $5$ ($\triangle$). Also
shown is the quenched potential at $\beta=3$ ($\bullet$), and the
broken string energy (dashed lines).} \label{fig:time}
\end{figure}

Using coarse lattices made it much easier to measure Wilson loops
$W(R,T)$ at propagation times $T$ large enough to accurately
identify the ground state energy at large $R$. Plots of the
time-dependent effective potential $V(R,T) = -\ln[W(R,T) /
W(R,T-1)]$ are given in Fig. \ref{fig:time}, and show that one
could easily be misled as to the shape of the potential if the
calculation is not done at large enough $T$, particularly at large
$R$.

In order to have some intuition for the length scales probed in
these simulations, it is instructive to use the physical value of
the string tension in four dimensions, $\sqrt\sigma=0.44$~GeV, to
in effect set the value of the dimensionful coupling constant in
QCD$_3$. Using the quenched continuum extrapolation $\beta a
\sqrt\sigma=1.33(1)$ \cite{Teper} one identifies, for example,
$a(\beta=3) \approx 0.2$~fm. Comparable, if slightly smaller,
estimates of the lattice spacing are obtained if simulation
results for the quenched or unquenched $\rho$ meson mass in
QCD$_3$ are identified with the physical mass.

This scale setting implies a string breaking distance $R_b \approx
1.5$~fm, which is numerically very similar to estimates of $R_b$
in four dimensions \cite{Sommer}. The results of Fig.
\ref{fig:time} suggest that propagation times somewhat in excess
of 1~fm are needed to adequately resolve string breaking in Wilson
loops on modestly coarse lattices; notice than an appreciable rise
in the potential remains even at $T \approx 0.6$~fm, which is
comparable to the longest propagation times resolved in the
four-dimensional simulations of Ref. \cite{CPPACS}.

Although longer propagation times $T$ are necessary to fully
isolate the ground state energy at the largest separations $R$
in Fig. \ref{fig:time}, these data nevertheless clearly support the
onset of string breaking. This follows from a comparison
of the unquenched and quenched potentials.
Of particular importance is the fact that any slope
that may exist in the unquenched potential data at large $R$
is clearly much smaller than the slope in the quenched data.
Such a small slope, if truly present, would imply a very large 
change in other physical quantities due to unquenching. In fact,
$m_\rho$ was found to change by less than 10\% with unquenching.
Hence the significant flattening of the unquenched potential data 
in Figs. \ref{fig:dpot} and \ref{fig:time}, compared to the 
quenched data, is very strongly indicative of the onset of 
string breaking. 

To summarize, string breaking due to dynamical fermions was
observed in three-dimensional lattice QCD. The results have clear
implications for current work on the four dimensional theory. In
particular, string breaking was readily observed using only Wilson
loop operators. The use of improved actions on coarse lattices
provided a crucial advantage in accessing the quark separations
and propagation times at which string breaking occurs. The use of
comparably coarse lattices in four dimensions would result in an
enormous improvement in computational efficiency, and could enable
string breaking to be demonstrated in realistic QCD.

I am indebted to Peter Lepage for several insightful discussions.
I also thank R. Woloshyn, R. Fiebig, M. Alford, and M. Teper for
fruitful conversations. This work was supported in part by the
Natural Sciences and Engineering Research Council of Canada.

\begin{references}

\bibitem[*]{Permanent} Permanent address.

\bibitem{Kuti}For a recent review see, e.g.,
J. Kuti, Report No. hep-lat/9811021.

\bibitem{Michael}See C. Michael, Report No. hep-ph/9710249,
and references therein.

\bibitem{CPPACS} S. Aoki {\it et al.}, CP-PACS collaboration,
Report No. hep-lat/9809185.

\bibitem{UKQCD} M. Talevi {\it et al.}, UKQCD collaboration,
Report No. hep-lat/9809182.

\bibitem{SESAMTchiL} G. Bali {\it et al.}, SESAM and T$\chi$L
collaborations, Nucl. Phys. B (Proc. Suppl.) {\bf 63}, 209 (1998).

\bibitem{DeTar} String breaking due to dynamical fermions at
nonzero temperature has been recently been reported in C. DeTar
{\it et al.}, Report No. hep-lat/9808028.

\bibitem{Gusken} S. G\"usken,  Nucl. Phys. B (Proc. Suppl.)
{\bf 63}, 16 (1998).

\bibitem{Drummond} I. T. Drummond, Phys. Lett. B {\bf 434},
92 (1998).

\bibitem{Wittig} O. Philipsen and H. Wittig,
Phys. Rev. Lett. {\bf 81}, 4056 (1998).

\bibitem{Koniuk} F. Knechtli and R. Sommer,
Phys. Lett. B {\bf 440}, 345 (1998); 
C. Stewart and R. Koniuk, hep-lat/9811012.

\bibitem{Teper} M. Teper, Phys. Rev. D {\bf 59}, 014512 (1999).

\bibitem{Kogut} E. Dagotto, A. Kocic and J.B. Kogut,
Nucl. Phys. B {\bf 362}, 498 (1991).

\bibitem{Prelim} Some this work was reported in preliminary form
in H. D. Trottier, Report No. hep-lat/9809183.

\bibitem{Poulis} G. I. Poulis and H. D. Trottier,
Phys. Lett. B {\bf 400}, 358 (1997).

\bibitem{Sommer} C. Alexandrou {\it et al.},
Nucl. Phys. B {\bf 414}, 815 (1994).

\bibitem{Alford} See, e.g., M. Alford, T. R. Klassen, 
and G. P. Lepage, Phys. Rev. D {\bf 58}, 034503 (1998).

\bibitem{Symanzik} K. Symanzik, Nucl. Phys. B {\bf 226},
187 (1983); M. L\"uscher and P. Weisz, Comm. Math. Phys. {\bf 97},
59 (1985).

\bibitem{Gottlieb} S. Gottlieb {\it et al.}, Phys. Rev.
D {\bf 35}, 2531 (1987).

\bibitem{Lepage} This argument is due to G. P. Lepage
(private communication). 
See also G. D. Moore, Nucl. Phys. B {\bf 523}, 569 (1998).

\bibitem{LepMac} G.P. Lepage and P.B. Mackenzie,
Phys. Rev. D {\bf 48}, 2250 (1993).

\bibitem{Burden}
C. Burden and A.N. Burkitt, Europhys. Lett. {\bf 3}, 545 (1987).

\bibitem{Temp} Finite temperature effects should be negligible
on these lattices, where $1/(N_\tau a) \lesssim 0.2 T_c$ 
[using the quenched estimate of $T_c$ in 
M. Teper, Phys. Lett. B {\bf 313}, 417 (1993)].

\bibitem{ImpStagg} Simulations were also done with
some improvement of the staggered fermion action; see
Ref. \cite{Prelim}.

\bibitem{Fiebig} A. Mihaly {\it et al.}, Phys. Rev.
D {\bf 55}, 3077 (1997).


\end{references}
\end{document}